\documentclass{article}
\usepackage{microtype}
\usepackage{graphicx}
\usepackage{subfigure}
\usepackage{tablefootnote}
\usepackage{booktabs} 
\usepackage{hyperref}

\usepackage[accepted]{icml2019}

\icmltitlerunning{Multi-Scale Embedded CNN for Music Tagging (MsE-CNN)}

\begin{document}

\twocolumn[
\icmltitle{Multi-Scale Embedded CNN for Music Tagging (MsE-CNN)}




\begin{icmlauthorlist}
\icmlauthor{Nima Hamidi Ghalehjegh}{to,ed}
\icmlauthor{Mohsen Vahidzadeh}{to,goo}
\icmlauthor{Stephen Baek}{too}
\end{icmlauthorlist}

\icmlaffiliation{to}{Department of Computer Science, The University of Iowa, Iowa City, Iowa, USA}
\icmlaffiliation{ed}{School of Music, The University of Iowa, Iowa City, Iowa, USA}
\icmlaffiliation{goo}{Civil and Environmental Engineering, The University of Iowa, Iowa City, Iowa, USA}
\icmlaffiliation{too}{Industrial and Systems Engineering, The University of Iowa, Iowa City, Iowa, USA}

\icmlcorrespondingauthor{Nima Hamidi Ghalehjegh}{nima-hamidi@uiowa.edu}

\icmlkeywords{Music Tagging, CNN, Deep Learning}

\vskip 0.3in
]



\printAffiliationsAndNotice{} 

\begin{abstract}
Convolutional neural networks (CNN) recently gained notable attraction in a variety of machine learning tasks: including music classification and style tagging. In this work, we propose implementing intermediate connections to the CNN architecture to facilitate the transfer of multi-scale/level knowledge between different layers. Our novel model for music tagging shows significant improvement in comparison to the proposed approaches in the literature, due to its ability to carry low-level timbral features to the last layer.
\end{abstract}

\section{Introduction}
\label{submission}
Music classification and style tagging is a traditional problem in music information retrieval (MIR) which entails predicting specific tags of a song, including genre, emotion, era, and instrumentation. Two main types of neural networkss used in music classification algorithms include convolutional neural networks (CNN) and recurrent neural networks (RNN), both of which were initially designed to facilitate different problems in image and language processing. However, the techniques mentioned above proved to be robust enough to be transferred to other applications successfully such as music tagging.\\
State-of-the-art computer vision algorithms use a technique called multi-scale feature pyramid network (FPN), which is capable of carrying low-level features from the lower layer to the higher layers in a neural networks. Concatenating these multi-scale features leads to improved accuracy \cite{Kirillov2019, lin2017feature, Selim2018, Kaiming2017}. Furthermore, U-Net is another novel architecture for image segmentation that consists of both a contracting path to capture context and a symmetric expanding path that enables precise localization. Having Mel-spectrum as an input to a CNN, a music classification is similar to the image classification task. We propose to use a multi-scale feature extraction structure in music classification, and we show that it can improve the accuracy of the model.\\
Many music tagging algorithms use spectrograms as input \cite{choi2016automatic, pons2017timbre, pons2017designing, choi2017convolutional}, while some others use raw audio files to extract relevant features on their own \cite{dieleman2014end,lee2017sample,zhu2016learning}. The successful performance of CNN in computer vision tasks led audio researchers to combine the use of Mel-spectrogram and CNN. Dieleman et al. compared the use of spectrograms and raw audio and reported a better performance in spectrograms \cite{dieleman2014end}. In this work, we focus on spectrograms as input to CNN and will build a new model upon currently available architectures to achieve improved performance.

\section{Development of the architecture}
In this study, we developed the architecture based on fully convolutional networks (FCN) as proposed in \cite{choi2016automatic}. We aim to improve their model by adding intermediate connections to the structure. In our proposed model, we add links in between every two convolutional layers in the FCN architecture, which allows transferring the features from these layers to deeper layers to combine low-level features with high-level features.

\subsection{Fully convolutional network}
Choi et. al. \cite{choi2016automatic} proposed FCNs with a different number of layers for music tagging and Figure 1 shows a five-layer FCN (FCN-5) version of their model. We selected this version because it showed the best results when applied to both MagnaTagATune and Million Song datasets. It includes five convolutional and max-pooling layers. The input to this model is a Mel-spectrogram of size $96\times1366\times1$, and more details are shown in Figure 1. Pooling size increases at each level, to increase the receptive filed gradually.\footnote{\url{https://github.com/keunwoochoi/music-auto_tagging-keras}}

\subsection{Multi-scale Embedded CNN (MsE-CNN)}
We propose that, by utilizing intermediate connections in our CNN architecture, we can carry important multi-scale features to the last layer for improved classification. We claim that such an approach will allow the model to learn low-level features such as musical texture and timbre as well as high-level temporal characteristics to improve the classification.\\
Figure 2 demonstrates our proposed architecture. At each level, we preserve the features before passing them through a convolutional layer, then concatenate them with results of the convolution by passing them through a max pooling operation. Therefore, at the output level, we have almost eight times as many features as FCN-5, while the increase in time complexity is negligible.\\

\begin{figure}[!htb]
\vskip -0.2in
\begin{center}
\centerline{\includegraphics[width=0.9\columnwidth]{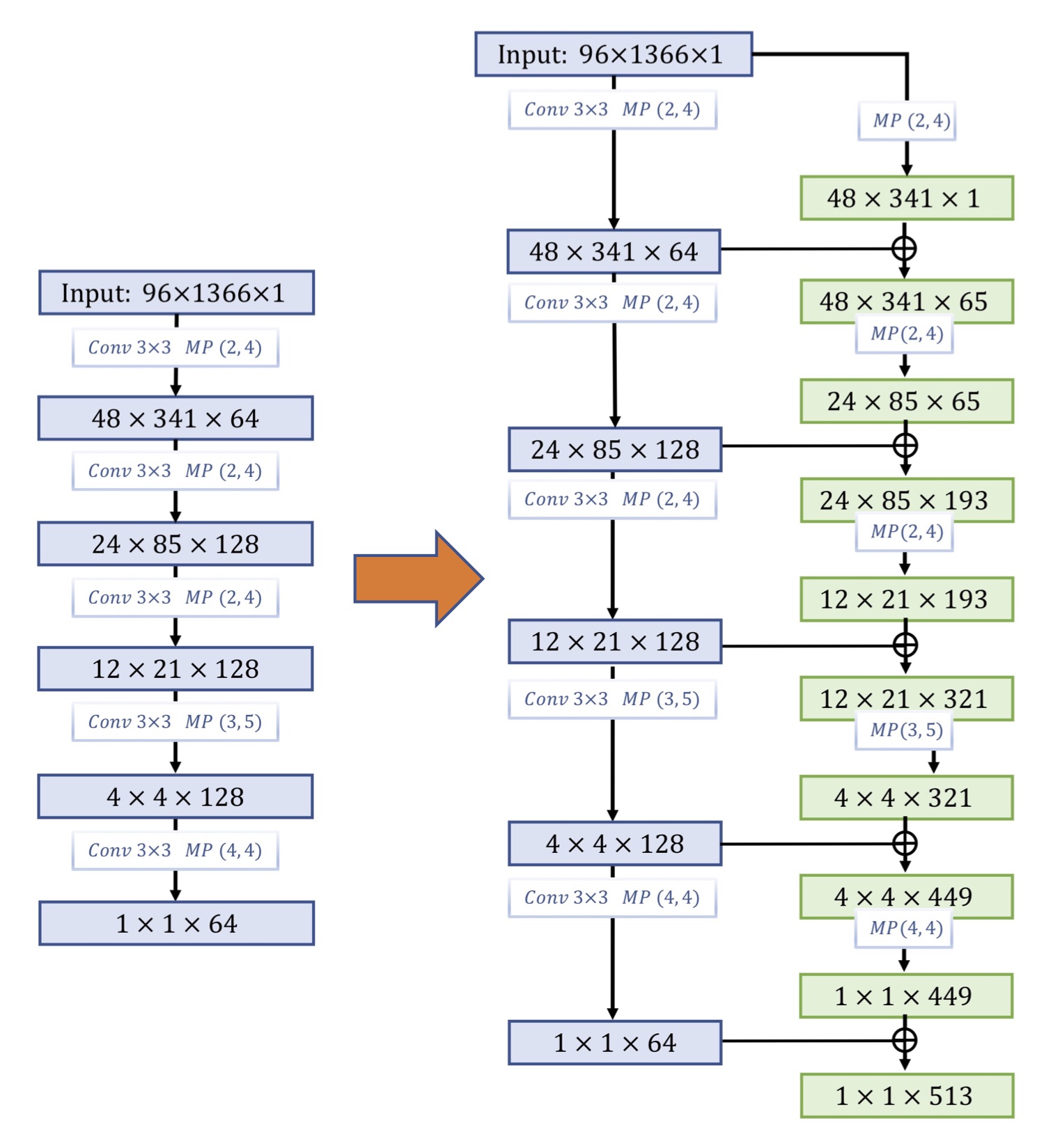}}
\caption{ A block diagram of the 5-layer CNN architecture proposed by Choi et. al. vs Multi-scale embedded Convolutional neural networks (MsE-CNN) for music classification. }
\label{icml-historical}
\end{center}
\vskip -0.2in
\end{figure}
\section{Experiment description}
In this study we used MagnaTagTune (MTT) dataset \cite{law2009evaluation} to tag songs and evaluate the performance of the model using both Area Under the Receiver Operating Characteristics (ROC-AUC) and Area Under the Precision-Recall Curve (PR-AUC). These methods have been used in literature extensively \cite{nam2015deep, dieleman2014end, van2014transfer, dieleman2013multiscale, hamel2011temporal} and allow us to perform a fair comparison of our model, MsE-CNN, with other architectures. 

\begin{table}[t]
\caption{Tagging accuracy of the proposed architectures (MsE-CNN) compared to previous results}
\label{Tagging Accuracies}
\vskip 0.15in
\begin{center}
\begin{small}
\begin{sc}
\begin{tabular}{lcr}
\toprule
Model & ROC-AUC & PR-AUC \\
\midrule
MsE-CNN (ours) & 0.914 & 0.423 \\
FCN-5 (reproduced) & 0.897 & 0.404 \\
FCN-4 \tablefootnote{Taken from \cite{choi2016automatic,pons2017end,pons2017timbre,nam2015deep,dieleman2014end,van2014transfer,dieleman2013multiscale,hamel2011temporal}. \label{xxx}}\textsuperscript{,}\tablefootnote{The PR-AUC results are based on reproduced version of the algorithm in \cite{pons2017end}. \label{xx}}   & 0.894 & 0.376 \\
End to End Learning \textsuperscript{\ref{xxx}}
 & 0.904 & 0.381 \\
Timbre CNN \textsuperscript{\ref{xxx},}\textsuperscript{\ref{xx}} & 0.893 & 0.349 \\
Bag of features and RBM \textsuperscript{\ref{xxx}} & 0.888 & -\\
1D convolutions \textsuperscript{\ref{xxx}} & 0.882 & -\\
Transferred learning \textsuperscript{\ref{xxx}} & 0.88 & -\\
Multi-scale approach \textsuperscript{\ref{xxx}} & 0.898 & -\\
Pooling MFCC \textsuperscript{\ref{xxx}} & 0.861 & -\\
\bottomrule
\end{tabular}
\end{sc}
\end{small}
\end{center}
\vskip -0.1in
\end{table}

\section{Discussion}

Using spectral representation of raw audio for music tagging shows significant improvement while utilizing CNN models. However, due to the complexity of timbral characteristics of sounds changing over time, the audio classification is considered different from image classification task. The change of timbral quality over time affects our perception of music, which is critical to the understanding of the genre, mood, instrumentation. We propose that in spectral representation of music, timbre is equivalent to texture and color in an image while long-term temporal structures are equivalent to shapes such as eye, hand, and nose. Similar to a vision task, learning musical textures and timbre as well as long-term characteristics is crucial in audio classification.\\
We believe a traditional CNN model learns long-term structures; however, it forgets timbral features in the final classification step. Similar to the intermediate connection in U-Net and FPN, one can improve music classification by transferring low-level characteristics via similar connections. Our model, MsE-CNN, is an experiment to support the idea that timbral fluctuation over time is disregarded while going deeper in a CNN model. Due to the larger receptive field in later layers in a CNN, the model starts to forget low-level features that carry a lot of textual details which are indeed crucial for audio classification.\\
Our experiment is more of a feasibility test to support our assumption; hence, it can be studied in greater details to find the optimal architecture. Indeed, such an approach can be extended to a variety of applications such as timbre analysis, instrument classification, audio clustering, and automatic music structure segmentation. In our future work, we aim to address a variety of MIR tasks to propose an optimal framework for the proposed training procedure. 
\bibliography{References}
\bibliographystyle{icml2019}
\end{document}